\documentclass[prl,twocolumn,floatfix,superscriptaddress,nofootinbib,nobalancelastpage]{revtex4-1}
\usepackage{dcolumn,amsmath}
\usepackage{graphicx}
\usepackage{color}
\usepackage{bm}
\usepackage{hyperref} 
\usepackage{physics,textcomp}
\usepackage{changepage}
\usepackage{lipsum}
\usepackage{subcaption}
\usepackage{natbib}
\usepackage{array}

\newcommand{\Ti}{$\mathcal{T}$}
\newcommand{\Par}{$\mathcal{P}$}
\newcommand{\eEDM}{{\em e}EDM}

\begin{document}
\title{$\mathcal{P}$,$\mathcal{T}$-odd energy shifts of the $^{173}$YbOH}

\author{Igor Kurchavov}  
\affiliation{Petersburg Nuclear Physics Institute named by B.P. Konstantinov of National Research Centre
"Kurchatov Institute", Gatchina 188300, Russia}
\author {Alexander Petrov}\email{petrov\_an@pnpi.nrcki.ru}
%
\affiliation{Petersburg Nuclear Physics Institute named by B.P. Konstantinov of National Research Centre
"Kurchatov Institute", Gatchina 188300, Russia}

\affiliation{St. Petersburg State University, St. Petersburg 199034, Russia} 

\date{Received: date / Revised version: date}
%
\begin{abstract}{
The energy shift in molecular spectra due to interaction of nuclear magnetic quadrupole moment ($M$) with electrons is equal to 
$\delta E_M = MW_M   P_{ M}$
where $W_M$ is a constant determined by the electronic structure of the molecule and $P_{ M}$ is a dimensionless constant. We extended the method for calculation of parity nonconservation effects in triatomic molecules developed in Ref. [A. Petrov and A. Zakharova, Phys. Rev. A {\bf 105}, L050801 (2022)] to the case of $P_{ M}$ constant
and applied it to $^{173}$YbOH in the first excited $v=1$ bending mode. 
Results of our calculations are required for extracting of the $M$ value from the YbOH experiment.

} 
\end{abstract}
\maketitle
\section{Introduction}
A search for the electron electric dipole moment (\eEDM), nuclear magnetic quadrupole moments (MQM), and other \Par,\Ti-odd (\Par\ is the space parity, \Ti\ is the time reversal) properties remains one of the key tests of the Standard Model of electroweak interactions and its most important extensions \cite{Fukuyama2012,PospelovRitz2014,YamaguchiYamanaka2020,YamaguchiYamanaka2021}. 
The best constraints on \eEDM\ were obtained in experiments with diatomic molecules YbF \cite{Hudson2011}, HfF$^+$ \cite{Cornell:2017} and ThO \cite{ACME:18}. 

The statistical accuracy of experiments on heavy-atom molecules may be significantly (by several orders of magnitude) enhanced by cooling and trapping the molecules in magneto-optical and/or dipole traps due to increased coherence time. Also the analysis of systematics is of utmost importance since it is what finally determines the accuracy limit of the measurement. 
 Previously it was shown  that due to existence of $\Omega$-doublet levels the experiments for searching of the $\mathcal{P}$,$\mathcal{T}$-odd effects on the ThO \cite{ACME:18,DeMille:2001,Petrov:14,Vutha:2010,Petrov:15,Petrov:17} 
or the HfF$^{+}$ \cite{Cornell:2017,Petrov:18} are very robust against a number of systematics.  Now it is clear that a large part of the great success with the HfF$^+$ and ThO  is due to its $\Omega$-doubling structure. Therefore, further success in the search for the \eEDM\ is also associated with the use of laser-cooled linear triatomic molecules like YbOH, RaOH \cite{Isaev:16, Isaev_2017, Kozyryev:17}. In case of the triatomic molecules, the role of $\Omega$-doublets used in the diatomic (like ThO and HfF$^+$) molecular experiments is taken over by the $l$-doublets of the excited $v = 1$ bending (distorting linear configuration) vibrational modes. The main advantage of YbOH and RaOH as compared to YbF and RaF is that the former due to bending vibrational structure has close levels of opposite parity $l$-doublets.

Any \eEDM\ experiment searches for an \eEDM\footnote{Together with \eEDM\ one always needs to consider scalar T,P-odd
electron-nuclear interaction, since its influence on the spectrum of molecules is identical to eEDM. For brevity, only the influence of \eEDM\ is mentioned in the present paper.} induced Stark shift
\begin{equation}
\delta E_e = d_e E_{\rm eff}   P,
\label{split}
\end{equation}
where $d_e$ is the value of the electron electric dipole moment; $E_{\rm eff}$ is the {\it effective electric field} acting on the electron in the molecule, which is the subject of molecular calculations
\cite{denis2019enhancement,prasannaa2019enhanced,gaul2020ab,Zakharova:21a, Zakharova:21b},
and $P$ is the polarization of the molecule by the external electric field.
 In our Letter \cite{PhysRevA.105.L050801} we have shown that the $l-doubling$ structure,
is, in general, different from $\Omega-doubling$, and the $P$ value tends to approach half of the maximum value for molecules with Hund's case $b$
and in particular for the YbOH molecule.

In turn nuclei with spin $I > 1/2$ gain a \Par,\Ti-odd-induced MQM that similarly to \eEDM\ interacts with an unpaired electron spin.
Measuring MQMs with the use  of molecules may provide improved limits on the strength of \Par,\Ti-odd nuclear forces, on the proton, neutron, and quark EDMs, on the quark chromo-EDMs, and on the QCD $\theta$ term and CP-violating quark interactions \cite{flambaum2014time, PhysRevLett.113.263006}. 

The scheme of measuring MQMs with the use of molecules is essentially the same as for the \eEDM. The energy shift in molecular spectra is equal to 
\begin{equation}
\delta E_M = MW_M   P_{ M},
\label{Msplit}
\end{equation}
where $M$ is the value of the MQM, $W_M$ (similarly to $E_{\rm eff}$) is determined by the electronic structure of the molecule and $P_{ M}$ (similarly to $P$) is the dimensionless constant (polarization). The $W_M$ constant for $^{173}$YbOH was calculated in \cite{maison2019theoretical}.
To extract $M = \delta E_M / W_M P_{ M} $ from the measured
shift $\delta E_M $, one needs to know both $W_M$ and $P_{ M}$ values.
Strictly speaking,  nuclei with spin $I \geq 1/2$ gain a Schiff moment that similarly to MQM contributes to nuclear spin dependent \Par,\Ti-odd energy shift.  In the current paper we consider only the MQM which is enhanced in the deformed nucleus of $^{173}$Yb \cite{maison2019theoretical}.

The YbOH molecule is a promising system for experiments looking for nonconservation effects such as \eEDM\ and nuclear MQM \cite{Kozyryev:17, PhysRevA.105.L050801}. 
For \eEDM\ search experiments, the spinless common $^{174}$Yb isotope would be ideal, while for MQM searches, the $^{173}$Yb isotope with nuclear spin $I=5/2$ has to be used. In Ref. \cite{PhysRevA.105.L050801} we have developed the method for calculation of the polarization $P$. The aim of the present paper is to extend the method to $P_{ M}$ and apply it to the ground rotational level of the first excited $v = 1$ bending vibrational mode of the $^{173}$YbOH molecule. Previously $P_M$ was calculated in diatomics \cite{Petrov:18b,Kurchavov:2020,Kurchavov2021}.

Since both \eEDM\ and MQM contribute to the measured energy splitting of the molecular $^{173}$YbOH spectra one needs to find a way to distinguish these two contributions. It is possible due to the different dependence of $P$ and $P_{ M}$ ($E_{\rm eff}$ and $W_M$ are the same for all hyperfine levels) on hyperfine level of the molecule. Then, performing the measurements on two (at least) different hyperfine levels (provided $P$ and $P_{ M}$ are known) allows one to distinguish between the \eEDM\ and MQM contributions. Therefore calculation of the $P$ for $^{173}$YbOH is also performed in the paper.  
\section{Method}
For the purpose of the present paper we present our Hamiltonian as
\begin{equation}
{\rm \bf\hat{H}} = {\rm \bf\hat{H}}_{\rm mol} + {\rm \bf\hat{H}}_{\rm hfs} + {\rm \bf\hat{H}}_{\rm ext},
\label{Hamtot}
\end{equation} 
where
${\rm \bf\hat{H}}_{\rm mol}$ is the molecular Hamiltonian as it is described in Ref. \cite{PhysRevA.105.L050801},
\begin{equation}
\begin{aligned}
 {\rm \bf\hat{H}}_{\rm hfs} = { g}_{\rm H} {\bf \rm I^H} \sum_a\left(\frac{\bm{\alpha}_{a}\times \bm{r}_{2a}}{r_{2a}^3 }\right) +
{ g}_{\rm Yb}{\mu_{N}} {\bf \rm I^{Yb}} \sum_a\left(\frac{\bm{\alpha}_a\times \bm{r}_{1a}}{{r_{1a}}^3}\right) \\
-e^2 \sum_q (-1)^q \hat{Q}^2_q({\bf \rm I^{\rm Yb}}) \sum_a \sqrt{\frac{2\pi}{5}}\frac {Y_{2q}(\theta_{1a}, \phi_{1a})}{{r_{1a}}^3}
\end{aligned}
\end{equation}
is the hyperfine interaction of electrons with Yb and H nuclei,
${ g}_{\rm Yb}$ and ${ g}_{\rm H}$ are the
 g factors of the ytterbium and hydrogen nuclei, $\bm{\alpha}_a$
 are the Dirac matrices for the $a$-th electron, $\bm{r}_{1a}$ and $\bm{r}_{2a}$ are their
 radius-vectors in the coordinate system centered on the Yb and H nuclei,
 $\hat{Q}^2_q({\bf \rm I^{\rm Yb}})$ is the quadrupole moment operator for the $^{173}$Yb nucleus,
index $a$ enumerates (as in all equations below) electrons of YbOH.
\begin{equation}
 {\rm \bf\hat{H}}_{\rm ext} =   -{ {\bf D}} \cdot {\bf E}
\end{equation}
describes the interaction of the molecule with the external electric field, and
{\bf D} is the dipole moment operator.

Wavefunctions were obtained by numerical diagonalization of the Hamiltonian (\ref{Hamtot})
over the basis set of the electronic-rotational-vibrational wavefunctions
\begin{equation}
 \Psi_{\Omega m\omega}P_{lm}(\theta)\Theta^{J}_{M_J,\omega}(\alpha,\beta)U^{\rm H}_{M^{\rm H}_I}U^{\rm Yb}_{M^{\rm Yb}_I}.
\label{basis}
\end{equation}
Here 
 $\Theta^{J}_{M_J,\omega}(\alpha,\beta)=\sqrt{(2J+1)/{4\pi}}D^{J}_{M_J,\omega}(\alpha,\beta,\gamma=0)$ is the rotational wavefunction, $U^{\rm H}_{M^{\rm H}_I}$ and $U^{\rm Yb}_{M^{\rm Yb}_I}$ are the hydrogen and ytterbium nuclear spin wavefunctions, $M_J$ is the projection of the molecular (electronic-rotational-vibrational) angular momentum $\hat{\bf J}$ on the laboratory axis, 
 $\omega$ is the projection of the same momentum on the $z$ axis of the molecular frame,
 $M^{\rm H}_I$ and $M^{\rm Yb}_I$ are the projections of the nuclear angular 
momenta of hydrogen and ytterbium on the laboratory axis, $P_{lm}(\theta)$ is the associated Legendre polynomial, and $\Psi_{\Omega m\omega}$ is the electronic wavefunction (see Ref. \cite{PhysRevA.105.L050801} for details).

In this  calculation functions with $\omega-m=\Omega=\pm1/2$, $l=0-30$, $m=0,\pm 1, \pm 2$ and $J=1/2,3/2,5/2$ (as in Ref. \cite{PhysRevA.105.L050801}) were included to the basis set (\ref{basis}).
Note, that the ground mode $v=0$ corresponds to $m=0$,
the first excited bending mode $v=1$ corresponds to $m=\pm 1$, the second excited bending mode has states with $m=0, \pm2$, etc.
 \begin{figure}[b]
    \includegraphics[width=0.5\textwidth]{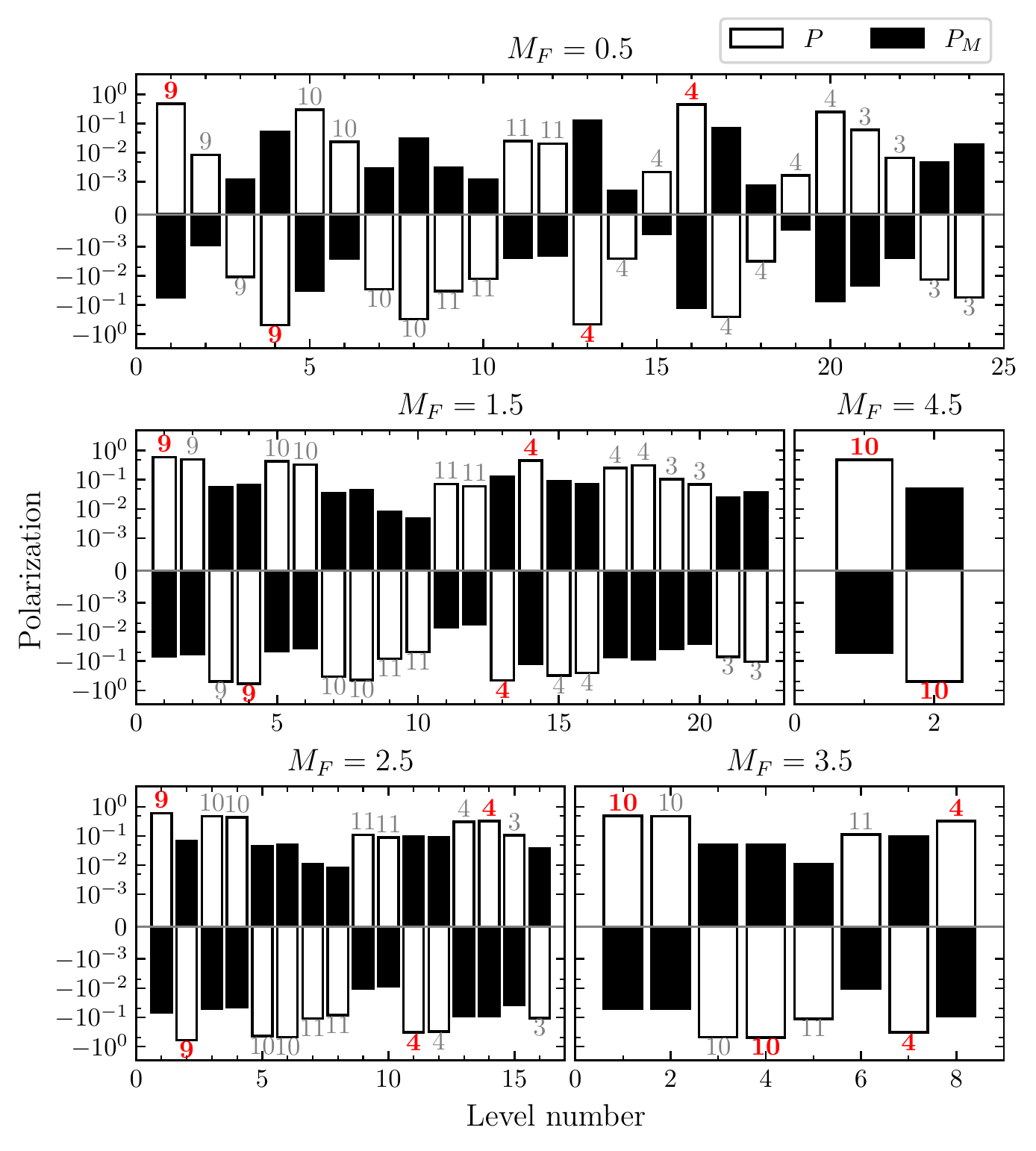}
    \caption{(Color online) The calculated polarization $P$, $P_M$ and ratio $P/P_M$. The abscissa numbering the states by increasing energy.
    Calculations have performed for electric field $E=150$ V/cm. For levels suggested for experiment the $P/P_M$ values are marked by red (bold style). }
    \label{EDMMQMshift}
\end{figure}
Provided that the {\it electronic-vibrational} matrix elements are known, the matrix elements of ${\rm \bf\hat{H}}$ between states in the basis set (\ref{basis}) can be calculated with help of angular momentum algebra \cite{LL77, PhysRevA.105.L050801} in the same way as for the diatomic molecules \cite{Petrov:11}.
Matrix elements required to calculate ${\rm \bf\hat{H}}_{\rm mol}$,  ${\rm \bf\hat{H}}_{\rm ext}$, and hyperfine interaction associated with the hydrogen nucleus were taken from Ref. \cite{PhysRevA.105.L050801}.

Matrix elements required to calculate hyperfine interaction with the ytterbium nucleus are
\begin{equation}
\begin{aligned}
A_{ \parallel} &= \frac{g_{\rm Yb}}{\Omega}
   \langle
   \Psi_{\Omega m\omega}P_{lm} |\sum_a\left(\frac{\bm{\alpha}_{a}\times
\bm{r}_{1a}}{r_{1a}^3}\right)
_z|\Psi_{\Omega m \omega}P_{l'm}\rangle \\
&= -1929~\delta_{ll'} {~ \rm MHz},
\end{aligned}
\end{equation}
\begin{multline}
A_{\perp} = {g_{\rm Yb}}\times \\
   \langle
   \Psi_{\Omega=1/2m\omega}P_{lm} |\sum_a\left(\frac{\bm{\alpha}_a\times
\bm{r}_{1a}}{r_{1a}^3}\right)
_+|\Psi_{\Omega=-1/2 m \omega-1}P_{l'm}\rangle \\
= -1856~ \delta_{ll'} {~ \rm MHz},
\end{multline}
\begin{multline}
e^2Qq_0 =  \langle
   \Psi_{\Omega m\omega}P_{lm} | \\
   e^2 \sum_q (-1)^q \hat{Q}^2_q({\bf \rm I^{\rm Yb}}) \sum_a \sqrt{\frac{2\pi}{5}}\frac {Y_{2q}(\theta_{1a}, \phi_{1a})}{{r_{1a}}^3} \\
   |\Psi_{\Omega m \omega}P_{l'm}\rangle = 3319~ \delta_{ll'} {~ \rm MHz}
\end{multline}
from Ref. \cite{Pilgram:21}.
Following Ref. \cite{PhysRevA.105.L050801} we neglect
 the $\theta$ dependence of the above matrix elements.
\section{Results and discussions}
\begin{figure}[t]
\includegraphics[width=0.49\linewidth]{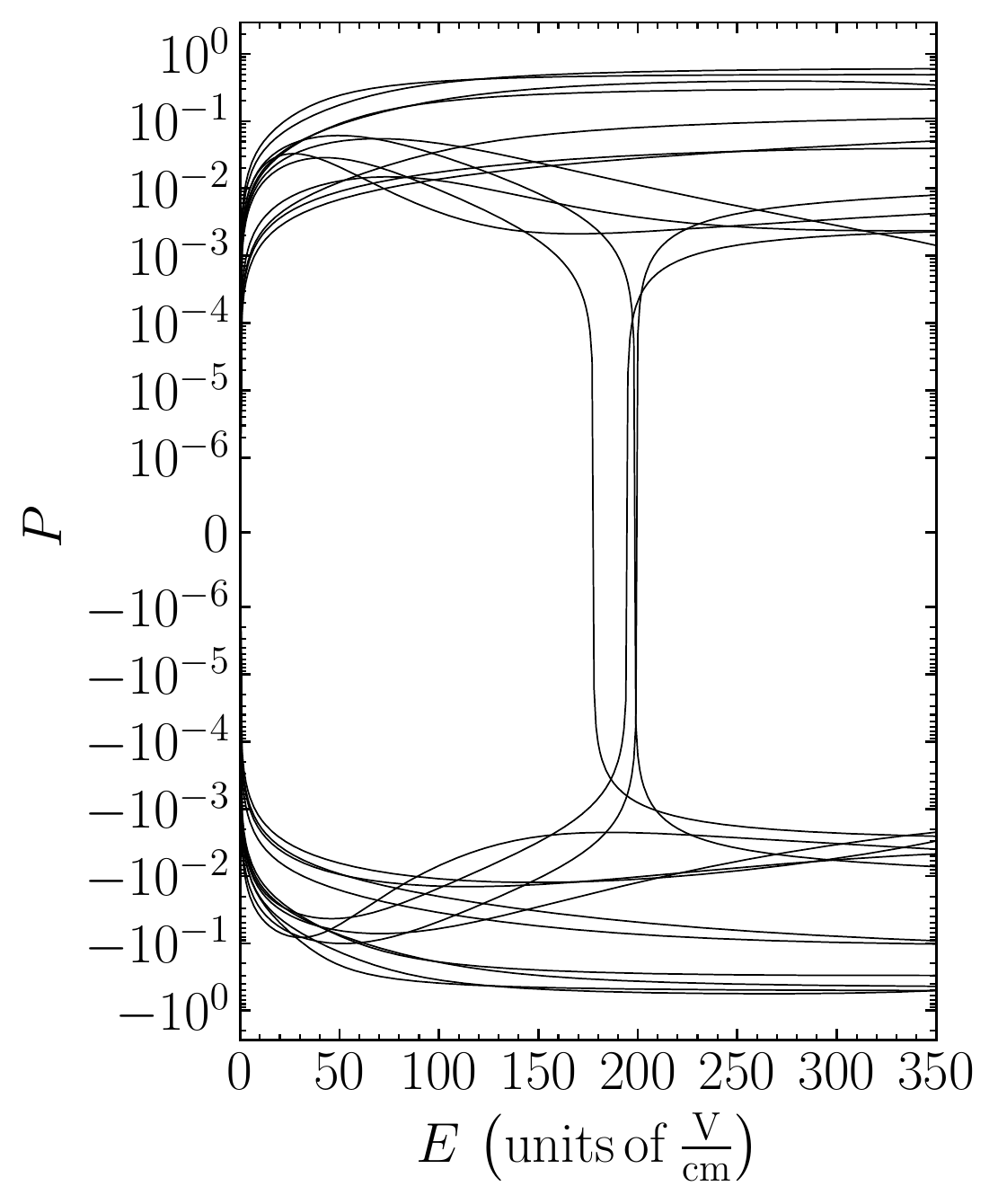}
\includegraphics[width=0.49\linewidth]{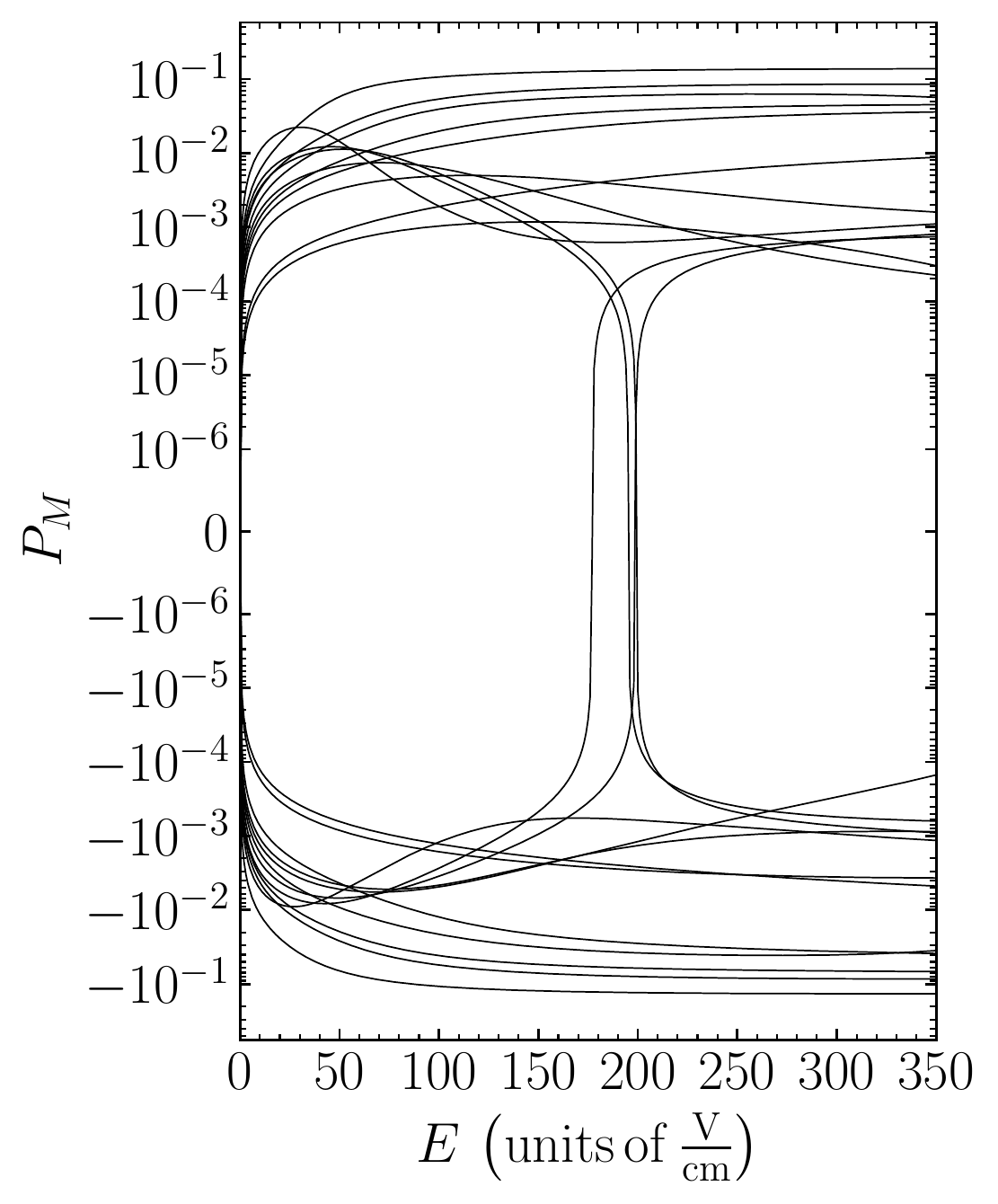}
\caption{\label{EDMMQMshift2} 
 Calculated polarizations $P$  and $P_M$ for the $M_F=0.5$ of the lowest $N=1$ rotational level
of the first excited the $v=1$ bending vibrational mode of $^{173}$YbOH as functions of the external electric field.}
\end{figure}
In Fig. \ref{EDMMQMshift} the calculated polarizations $P$, $P_M$, as well as ratio $P/P_M$ for the lowest $N=1$ rotational level
of the first excited $v=1$ bending vibrational mode of the $^{173}$YbOH for the external electric field $E=150$ V/cm are presented as a bar chart for clarity. Within the $M_F$ manifolds the levels are ordered
by the energy value.
Here $M_F=M_J+{M^{\rm H}_I}+{M^{\rm Yb}_I}$ is the projection of the total molecular (electronic-rotational-vibrational-nuclear spins) angular momentum ${\bf F}$ on the laboratory axis. Numerical data for Fig. \ref{EDMMQMshift} are given in Table S1 in supplementary.
There are 24 levels for $M_F=1/2$, 22 levels for $M_F=3/2$, 16 levels for $M_F=5/2$, 8 levels for $M_F=7/2$ and 2 levels for $M_F=9/2$.
As an example, in Fig. \ref{EDMMQMshift2} the calculated $P$ and $P_M$
for $M_F=1/2$ as functions of the external electric field are presented.
Electric field $E=150$ V/cm ensures almost saturated values for $P$ and $P_M$.
Calculations showed that all levels have polarizations $P<0.65$ and $P_M<0.15$.
Energy levels for all $M_F$ values as functions of the external electric field are presented
in Fig. \ref{Energy}. Numerical data for Fig. \ref{Energy} are given in Table S2 in supplementary.

Since MQM induced energy shift is proportional to $P_M$ (see Eq. (\ref{Msplit})) for MQM searches the levels with large $P_M$ values are preferred. Beyond this,  
 to distinguish \eEDM\ and MQM contributions, the levels with different $P/P_M$ ratios have to be used. For the levels satisfying these conditions the $P/P_M$ values are marked by red (bold style) in Fig. \ref{EDMMQMshift}. In Fig. \ref{Energy} the corresponding levels are marked by the bold style. These levels are the first, fourth, thirteenth, and sixteenth for $M_F=0.5$; the first, fourth, thirteenth, and fourteenth for $M_F=1.5$;  the first, second, eleventh, and fourteenth for $M_F=2.5$; the first, fourth, seventh, eighth for $M_F=3.5$; and the first and second for $M_F=4.5$ (the only levels for $M_F=4.5$).
 
 As an example of the proposed \eEDM\ contribution exclusion scheme, let us consider the first and sixteenth levels of $M_F=0.5$.
 For the first level we have $P=0.4833$, $P_M=-0.0542$, and $|P/P_M|=8.91$. Then the energy shift induced by \eEDM\ and MQM is $\delta E^1 = \delta E^1_e + \delta E^1_M  = 0.4833d_e E_{\rm eff} - 0.0542MW_M$. For the sixteenth level we have $P=0.4515$, $P_M=-0.1235$, and $|P/P_M|=3.66$. The corresponding energy shift is $\delta E^{16} = \delta E^{16}_e + \delta E^{16}_M  = 0.4515d_e E_{\rm eff}  -0.1235MW_M$. 
 Then we have $\delta E^{1}  -1.07 \delta E^{16} \approx 0.078MW_M$. So, we managed to exclude the contribution from the \eEDM\ and can extract $M$ (our final goal, $W_M$ is known from Ref. \cite{maison2019theoretical}) from the measured  $\delta E^{1}$  and $\delta E^{16}$.  If the ratios $P/P_M$ for the considered levels  were the same we would not be able to write appropriate (depending only on $M$) linear combination of $\delta E^1$ and $\delta E^{16}$. Thus we should choose levels with different $P/P_M$ ratios. Formally, it is enough to select two levels (independently whether they have the same $M_F$ or not) with different $P/P_M$ ratios to extract the $M$ value. However, using three or more levels can help to improve statistics, check data (both experimental $\delta E$ and calculated $P$, $P_M$) for consistency. Using only two levels in the experiment may not be practical, as the disentanglement of \eEDM\ and MQM contributions is depending on the accuracy of the computed $P$ and $P_M$ values.
 We note also that our selection of the levels is only an example. On the base of Fig. \ref{EDMMQMshift} one can select another appropriate levels for the MQM search. 
 
Finally we calculated the polarizations $P$ and $P_M$ associated with \eEDM\ and MQM energy shifts for the
$^{173}$YbOH molecule in the first excited bending mode. The levels most suitable for the MQM search are determined.
\section{acknowledgement}
This work is supported by the Russian Science Foundation grant No. 18-12-00227.
\begin{figure}[ht!]
\includegraphics[width=0.95\linewidth]{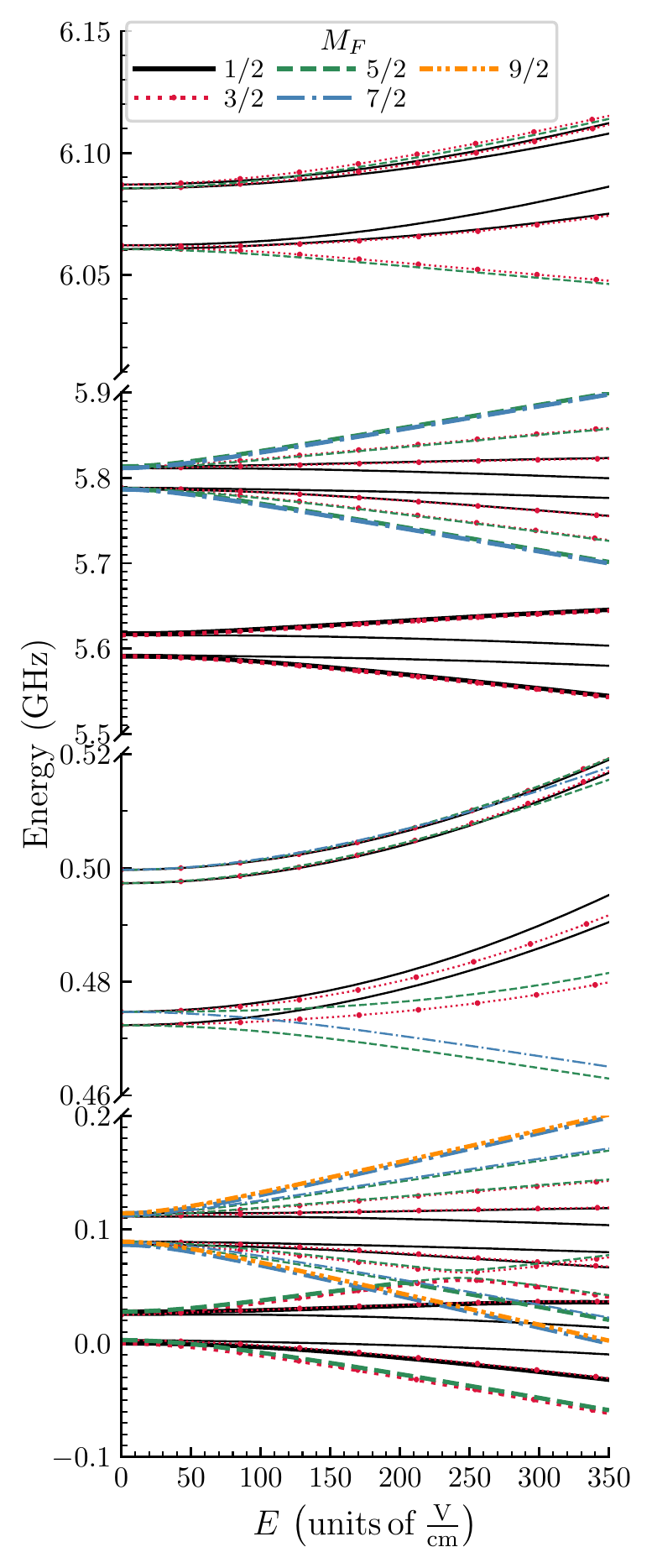}
\caption{\label{Energy} 
(Color online) The calculated energies of the lowest $N=1$ rotational 
of the first excited $v=1$ bending vibrational mode of $^{173}$YbOH as functions of the external electric field. 
Zero energy level corresponds to the the lowest energy of $N=1$ states
at zero electric field.
}
\end{figure}

\end{document}